# Heavy Quark Solitons: Towards Realistic Masses

J. Schechter[a], A. Subbaraman[b], S. Vaidya[a] and H. Weigel[c]

[a]Department of Physics, Syracuse University
Syracuse, NY 13244–1130, USA

[b]Department of Physics, University of California
Irvine, CA 92717, USA

[c]Institute for Theoretical Physics, Tübingen University
D-72076 Tübingen, Germany

**ABSTRACT**

A generalization of the effective meson Lagrangian possessing the heavy quark symmetry to finite meson masses is employed to study the meson mass dependence of the spectrum of S– and P wave baryons containing one heavy quark or anti-quark. These baryons are described as respectively heavy mesons or anti-mesons bound in the background of a soliton, which is constructed from light mesons. No further approximation is made to solve the bound state equation. For special cases it is shown that the boundary conditions, which have to be satisfied by the bound state wave–functions and stem from the interaction with the light mesons, may impose additional constraints on the existence of bound states when finite masses are assumed. Two types of models supporting soliton solutions for the light mesons are considered: the Skyrme model of pseudoscalars only as well as an extension containing also light vector mesons. It is shown that only the Skyrme model with vector mesons provides a reasonable description of both light and heavy baryons. Kinematical corrections to the bound state equations are included in the discussion.

## 1. Introduction

There has been a good deal of recent interest in the study of the heavy baryons in the bound state picture [1, 2] with the assumption of heavy quark spin symmetry [3]. Many aspects of this interesting but technically involved problem have been treated by various groups [4]–[8]

The baryons under study have the schematic quark structure $qqQ$, where $q$ stands for a light quark and $Q$ for a heavy quark. In the given approach they are realized as heavy mesons $(\bar{q}Q)$ bound to light baryons $(qqq)$. The light baryons are treated as soliton excitations of a light meson Lagrangian. Then a piece is added to the chiral Lagrangian in order to describe the heavy mesons and their interactions with the light ones. The bound state equation is the equation of motion of the heavy meson field in the background field of the light soliton. This turns out to involve several coupled differential equations but, in the infinite meson mass, $M \to \infty$ limit it simplifies enormously [9] to the evaluation of the matrix elements of an operator of the form $a 1\!\!1 + b \sigma_i \tau_i$ in a suitable space. This simple result requires not only $M \to \infty$ but also that the light baryon mass be formally infinite.

The interest of this simple result attaches to the fact that it holds for a very large class of models in the combined large $N_c$ and large $M$ limit. However it is very desirable for the sake of comparison with experiment to understand the corrections due to using finite heavy meson mass and finite light baryon mass. For example, the orbitally excited heavy baryon states turn out unrealistically to be degenerate with the ground state in the simple limit. Of course once one goes away from the symmetry limit there are many options. In the present paper we shall deal with the simplest generalization of the heavy meson Lagrangian to finite heavy meson masses and shall solve the coupled differential equations *exactly* (numerically) for the states of interest. It turns out that the resulting equations are exactly of the same structure as the homogeneous part of those which arose in the "K-cranking" treatment [11, 12] of the $SU(3)$ Skyrme model with (light) vector mesons so that existing technology may be used.

Previously [9] it was shown how the coupled differential equations could be approximated by a single Schrödinger like equation. This suggested that the effect of finite light baryon mass could be estimated by replacing the heavy meson mass in this equation by the reduced mass. The conclusions of that approximate analysis were first that each of the finite heavy meson mass and finite light baryon mass corrections were very important and second that it was just about impossible to understand the existing experimental data with a light meson Lagrangian containing only pseudoscalar fields. It seemed that the latter problem could be solved if the Lagrangian also contained light vector mesons. Earlier [7] the coupled equations had been solved for the ground state using the light pseudoscalar only Lagrangian and approximating the time component of the heavy vector field by its leading order in $1/M$ piece. Later this approximation has been shown to be justified in that model [10]. The present analysis uses a different method in which it is unnecessary to make that approximation.

We confirm here that the light vector mesons seem to be very important to understand the existing experimental data. After fitting an unknown light vector-heavy meson coupling parameter to the binding energy of the $\Lambda_b(5641)$ baryon we are able to successfully predict the binding energy of the $\Lambda_c(2285)$ baryon as well as that of a recently observed [13] candidate for its first orbital excitation [14]. Furthermore we find that the approximate Schrödinger like equation mentioned above is accurate for the ground state but not reliable for the excited heavy baryon state in the charm sector. It seems useful to understand the accuracy of this equation when one recognizes that the parameters of the light -heavy meson interactions are still not conclusively established and that the light soliton models themselves may require important corrections [15]. An initial understanding of the results of changes or of extensions of the model may be more simply obtained with the Schrödinger like equation.

Amusingly, we find that the present model, including light vector mesons, appears to give quite a reasonable account of the "ordinary" hyperon binding energies.

We also investigate the so-called pentaquark states [8] in the present model. These are postulated states of the form $qqqq\bar{Q}$. In the present approach they arise as negative energy bound states of the heavy meson $(\bar{q}Q)$ in the background soliton field. We note that in the simple limit it is easier to work with their anti-particles which correspond to positive energy bound states of the heavy meson in the anti-soliton field. In any event it turns out that these states, which are slightly bound in the simple limit, become unbound in the model with light vectors both in the charm and bottom sectors.

In connection with our investigation of the penta states we found the interesting feature that a certain state which was bound in the infinite $M$ limit did not satisfy the appropriate boundary conditions near $r = 0$ for any finite value, no matter how large, of $M$. Thus it appears that the large $M$ limit and the $r \to 0$ limit (which is needed for obtaining the standard heavy limit results) do not necessarily commute with each other.

This paper is organized as follows. Section 2 contains a brief summary of the underlying chiral Lagrangian including light vector mesons. It also gives the *ansatz* for the light soliton and for the bound heavy mesons in the P-wave and S-wave orbital states. The latter are suitable for describing, respectively, the ground state baryon and its first orbitally excited state as well as the low lying penta quark states. In section 3 we obtain, from symmetry considerations, the wave–functions and binding energies of the above mentioned states in the heavy spin symmetry limit. In section 4 we discuss the boundary conditions at large and small r of the coupled differential equations (using, for simplicity, the model in which the light vector mesons are not present). The fact that the $M \to \infty$ and the $r \to 0$ do not necessarily commute is illustrated. In section 5 the detailed behavior of the wave–functions and binding energies of the low lying states is discussed for the model without light vectors. This is generalized to the model including light vectors in section 6. The behavior of the model as a function of an important parameter describing the heavy meson-omega meson coupling constant is treated. In section 7 the kinematical effects of finite light baryon mass are estimated and our final numerical predictions are presented. It should be noted that the differential equations themselves, their large $M$ limits and the method of solution are explicitly given in Appendix A. Section 8 contains a summary and discusses directions for further work.

## 2. Description of the Model

Following the bound state picture we regard the heavy baryon as a bound state of a heavy meson in the background field of a Skyrmion. In turn, the Skyrmion corresponds to a light baryon which arises as a soliton excitation of an effective Lagrangian constructed from the light pseudoscalar and light vector meson fields. For the sector of the model describing the light pseudoscalar and vector mesons we adopt the chirally invariant Lagrangian discussed in detail in the literature [16, 17]. This Lagrangian can be decomposed into a normal parity part

$$\mathcal{L}_S = f_\pi^2 \text{tr}\,[p_\mu p^\mu] + \frac{m_\pi^2 f_\pi^2}{2} \text{tr}\left[U + U^\dagger - 2\right] - \frac{1}{2} \text{tr}\left[F_{\mu\nu}(\rho) F^{\mu\nu}(\rho)\right] + m_V^2 \text{tr}\,[R_\mu R^\mu] \qquad (2.1)$$

and a part which contains the Levi-Cevita tensor, $\epsilon_{\mu\nu\alpha\beta}$. The action for the latter is most conveniently displayed using differential forms $p = p_\mu dx^\mu$, etc.

$$\begin{aligned}\Gamma_{\text{an}} &= \frac{2N_c}{15\pi^2} \int Tr(p^5) \\ &+ \int Tr\left[\frac{4i}{3}(\gamma_1 + \frac{3}{2}\gamma_2)Rp^3 - \frac{g}{2}\gamma_2 F(\rho)(pR - Rp) - 2ig^2(\gamma_2 + 2\gamma_3)R^3 p\right].\end{aligned} \qquad (2.2)$$

In eqs (2.1) and (2.2) we have introduced the abbreviations

$$p_\mu = \frac{i}{2}\left(\xi \partial_\mu \xi^\dagger - \xi^\dagger \partial_\mu \xi\right) \quad \text{and} \quad v_\mu = \frac{i}{2}\left(\xi \partial_\mu \xi^\dagger + \xi^\dagger \partial_\mu \xi\right) \qquad (2.3)$$

for the pseudovector and vector currents of the light pseudoscalar fields. Furthermore $\xi$ refers to a square root of the chiral field, i.e. $U = \xi^2$. Finally $F_{\mu\nu}(\rho) = \partial_\mu \rho_\nu - \partial_\nu \rho_\mu - ig[\rho_\mu, \rho_\nu]$ denotes the field tensor associated with the vector mesons $\rho$ and $\omega$, which are incorporated via $\rho_\mu = \left(\omega_\mu 1\!\!1 + \rho_\mu^a \tau^a\right)/2$ in the two flavor reduction. $R_\mu = \rho_\mu - v_\mu/g$ transforms simply under chiral transformations. The parameters $g, \gamma_1$, etc. can be determined (or at least constrained) from the study of decays of the light vector mesons such as $\rho \to 2\pi$ or $\omega \to 3\pi$. For details we refer to ref. [17]. The action for the light degrees of freedom ($\int \mathcal{L}_S + \Gamma_{\text{an}}$) contains static soliton solutions. The appropriate *ansätze* are

$$\xi(\boldsymbol{r}) = \exp\left(\frac{i}{2}\hat{\boldsymbol{r}}\cdot\boldsymbol{\tau}F(r)\right), \quad \omega_0(\boldsymbol{r}) = \frac{\omega(r)}{g} \quad \rho_{i,a}(\boldsymbol{r}) = \frac{G(r)}{gr}\epsilon_{ija}\hat{r}_j \qquad (2.4)$$

while all other field components vanish. These solutions have widely been employed to investigate static properties of the light baryons; see [18] for reviews.

We next require the part of the action which describes the chirally invariant coupling of the light pseudoscalar and vector mesons to their counterparts containing one heavy quark. In a suitable infinite heavy mass limit this part of the action has the additional heavy spin symmetry and the leading term in a $1/M$ expansion is unique – see for example eq.(3.24) of ref. [19]. A minimal extension to finite M of this action is given in eq.(3.25) of ref. [19] which we write out as:

$$\begin{aligned}\mathcal{L}_H &= D_\mu P \left(D^\mu P\right)^\dagger - \frac{1}{2} Q_{\mu\nu}\left(Q^{\mu\nu}\right)^\dagger - M^2 P P^\dagger + M^{*2} Q_\mu Q^{\mu\dagger} \\ &+ 2iMd\left(Pp_\mu Q^{\mu\dagger} - Q_\mu p^\mu P^\dagger\right) - \frac{d}{2}\epsilon^{\alpha\beta\mu\nu}\left[Q_{\nu\alpha}p_\mu Q_\beta^\dagger + Q_\beta p_\mu \left(Q_{\nu\alpha}\right)^\dagger\right] \\ &- \frac{2\sqrt{2}icM}{m_V}\left\{2Q_\mu F^{\mu\nu}(\rho) Q_\nu^\dagger + \frac{i}{M}\epsilon^{\alpha\beta\mu\nu}\left[D_\beta P F_{\mu\nu}(\rho) Q_\alpha^\dagger + Q_\alpha F_{\mu\nu}(\rho)(D_\beta P)^\dagger\right]\right\}.\end{aligned} \qquad (2.5)$$

Here we have allowed the mass $M$ of the heavy pseudoscalar meson $P$ to differ from the mass $M^*$ of the heavy vector meson $Q_\mu$. Note that the heavy mesons are conventionally defined as *row* vectors in isospin space. The covariant derivative introduces the additional parameter $\alpha$:

$$D_\mu P^\dagger = (\partial_\mu - i\alpha g\rho_\mu - i(1-\alpha)v_\mu) P^\dagger = (\partial_\mu - iv_\mu - ig\alpha R_\mu) P^\dagger \qquad (2.6)$$

for example, while the covariant field tensor of the heavy vector meson is defined as

$$(Q_{\mu\nu})^\dagger = D_\mu Q_\nu^\dagger - D_\nu Q_\mu^\dagger. \qquad (2.7)$$

It should be stressed that the assumption of infinitely large masses for the heavy mesons has not been made in (2.5). The coupling constants $d, c$ and $\alpha$ which appear have still not been very accurately determined. In particular there is no direct experimental evidence for the value of $\alpha$, which would be unity if a possible definition of light vector meson dominance for the electromagnetic form factors of the heavy mesons were to be adopted (see ref.[20]). We shall consider $\alpha$ as a parameter here. The other parameters in (2.5) will be taken to be:

$$\begin{aligned} d &= 0.53 \quad c = 1.60 \\ M &= 1865\,\text{MeV} \quad M^* = 2007\,\text{MeV} \qquad \text{D} - \text{meson} \\ M &= 5279\,\text{MeV} \quad M^* = 5325\,\text{MeV} \qquad \text{B} - \text{meson}. \end{aligned} \qquad (2.8)$$

¿From studies [1] in the bound state approach to the SU(3) Skyrme model we expect the ground state heavy baryon when the heavy meson is bound in an orbital P–wave while the first excited state is expected when the heavy meson is bound in an orbital S–wave. The apparent reversal from the usual expectation is due to the spin-isospin mixing in the Skyrme approach. In the context of the heavy quark symmetry it is, of course, necessary to also include the heavy vector meson fields. The corresponding *ansatz* for the P–wave

$$\begin{aligned} P^\dagger &= \frac{\Phi(r)}{\sqrt{4\pi}} \hat{\boldsymbol{r}} \cdot \boldsymbol{\tau} \chi e^{i\epsilon t}, \qquad Q_0^\dagger = \frac{\Psi_0(r)}{\sqrt{4\pi}} \chi e^{i\epsilon t}, \\ Q_i^\dagger &= \frac{1}{\sqrt{4\pi}} \left[ i\Psi_1(r)\hat{r}_i + \frac{1}{2}\Psi_2(r)\epsilon_{ijk}\hat{r}_j\tau_k \right] \chi e^{i\epsilon t} \end{aligned} \qquad (2.9)$$

defines four radial functions. $\chi$ represents a constant (iso) spinor. Hence $P^\dagger$ and $Q_\mu^\dagger$ represent isospinors. As a matter of fact these *ansätze* are identical to those used to compute induced strange fields in the framework of the collective approach to the Skyrme model with vector mesons [11, 12]. In that case $\chi$ parametrizes the angular velocities for rotations into strange directions. It should be noted that apart from $\chi$ the *ansätze* (2.9) carry zero grand spin and negative parity. The former is defined as the vector sum of total spin and isospin. As $\chi$ has to be interpreted as an isospinor, the total grand spin of the *ansatz* (2.9) is 1/2. Since the heavy mesons carry negative parity the resulting heavy baryon will have positive parity. For the *ansätze* in the S–wave channel

$$\begin{aligned} P^\dagger &= \frac{\Phi(r)}{\sqrt{4\pi}} \chi e^{i\epsilon t}, \qquad Q_0^\dagger = \frac{\Psi_0(r)}{\sqrt{4\pi}} \hat{\boldsymbol{r}} \cdot \boldsymbol{\tau} \chi e^{i\epsilon t}, \\ Q_i^\dagger &= \frac{i}{\sqrt{4\pi}} \left[ \Psi_1(r)\hat{\boldsymbol{r}} \cdot \boldsymbol{\tau} \hat{r}_i + \Psi_2(r)r\boldsymbol{\tau} \cdot \partial_i \hat{\boldsymbol{r}} \right] \chi e^{i\epsilon t} \end{aligned} \qquad (2.10)$$

the grand spin is still 1/2 while the parity is positive. Thus the resulting excited heavy baryon has negative parity. The phase conventions in (2.9, 2.10) guarantee real radial functions. It should be remarked that we actually are considering conjugate heavy meson fields. Hence the exponential reads $e^{i\epsilon t}$ rather than $e^{-i\epsilon t}$. In earlier treatments, the time component of the heavy vector field $Q_0$ had been eliminated using the heavy quark approximation

$$Q_0 \approx \frac{1}{M^{*2}} D^i \dot{Q}_i = \mathcal{O}\left(\frac{1}{M^*}\right). \qquad (2.11)$$

For the current investigation we will not need to make this approximation. The bound state equations may be obtained by substituting these *ansatze* in (2.5). After some computation this yields effective Lagrange densities for the radial functions $\Phi, \Psi_0, \Psi_1$ and $\Psi_2$. These are displayed in appendix A for both the S– and P wave channels. The leading pieces in the limit $M = M^* \to \infty$ are also presented in this appendix. Furthermore the method used to solve the bound state equations for finite masses is described.

## 3. Bound States in the Heavy Mass Limit

Before discussing the exact solutions to the bound state equations emerging from (2.5) for the S– and P wave heavy mesons it is illuminating to review the results associated with the heavy mass limit, *i.e.* $M \to \infty$ and $M^* \to \infty$. In that limit the wave–functions receive their only support at the origin $r = 0$. Hence the binding energy is given by the negative of the potential for the heavy mesons at the origin. Since this potential is generated by the static soliton the binding energy is extracted from the light meson profiles $F, G$ and $\omega$ at $r = 0$. For this purpose one substitutes the expansions

$$F(r) \approx -\pi + F'(0)r + \ldots, \quad G(r) \approx -2 + \frac{G''(0)}{2}r^2 + \ldots \quad \text{and} \quad \omega(r) \approx \omega(0) + \ldots \qquad (3.1)$$

of the profile functions for the light mesons into eqs (A.4) and (A.2). Note that $G$ and $\omega$ have been redefined compared to [9] ; see Appendix B. The binding energy is defined by $\epsilon_B = M - |\epsilon|$, *i.e.* $M^2 - \epsilon^2 \approx 2M\epsilon_B$. Here $\epsilon$ represents the value of the energy which leads to a regular solution of the bound state equations corresponding to (A.2) and (A.4). In what follows we will refer to this value of $\epsilon$ as the bound state energy. For the P wave the binding energy in the large $M$ limit results in [9]

$$\epsilon_B = \frac{3}{2}dF'(0) - \frac{3\sqrt{2}c}{gm_V}G''(0) + \frac{\alpha}{2}\omega(0). \qquad (3.2)$$

Simultaneously the radial functions in (2.9) are related via

$$\Psi_1 = -\Phi, \quad \Psi_2 = -2\Phi \qquad (P - \text{wave}). \qquad (3.3)$$

For the S wave the same binding energy (3.2) is obtained in the large $M$ limit, however, in this case the radial functions in (2.10) are related via

$$\Psi_1 = -\Phi, \quad \Psi_2 = \Phi \qquad (S - \text{wave}). \qquad (3.4)$$

The reason the binding energies are identical in the two channels is that the modifications arising in the differential equations when going from the $P-$ to the $S$ wave are subleading in the limit $M, M^* \to \infty$. Nevertheless, there may be significant differences in the binding energies of the S and P wave mesons when the masses are kept finite. The reason is that, due to the centrifugal barrier, the corresponding wave–functions are substantially different at $r \approx 0$. We will see later that the $P$ wave state is the ground state of the heavy baryon that results from the binding, while the $S$ wave state is interpreted as the first radially excited state. We have already argued above that the vicinity of $r \approx 0$ determines the binding energy. It will, of course, be of great interest to see how well the relations (3.2)-(3.4) are satisfied when finite masses are assumed. Technically speaking, the question is whether or not the limits $M = M^* \to \infty$ and $r \to 0$ commute.

In the limit $M = M^* \to \infty$ the structure of the wave–functions based on their symmetries has been worked out in ref. [9]. ¿From that scheme one may extract relations like (3.3) and (3.4) for other channels as well. The starting point is the heavy meson field $H$, which combines the heavy pseudoscalar and vector meson fields moving with a fixed four–velocity $V^\mu$

$$H^a = \frac{1}{2}\left(1 + \gamma_\mu V^\mu\right)\left(i\gamma_5 P'^a + \gamma^\nu Q'^a_\nu\right), \qquad \bar{H} = \gamma_0 H^\dagger \gamma_0 \ . \tag{3.5}$$

This parametrization makes the SU(2) spin symmetry transparent. The superscript $a$ labels the light flavor content. In the rest frame $H$ may be decomposed as

$$\bar{H}^a = \begin{pmatrix} 0 & 0 \\ \bar{H}^a_{lh} & 0 \end{pmatrix}, \quad (l, h = 1, 2). \tag{3.6}$$

The spatial representation of the heavy meson wave–function $\bar{H}^a_{lh}$ has been discussed in ref. [9] and reads[a]

$$\bar{H}^a_{lh} = \frac{u(r)}{\sqrt{2M}}\left(\hat{\boldsymbol{r}} \cdot \boldsymbol{\tau}\right)_{ad} \bar{\Psi}_{dl}\left(g, g_3\right) \chi'_h \ . \tag{3.7}$$

The radial part, $u(r)$, needs no specification other than that it be strongly peaked[b] in the vicinity of $r = 0$. Here $g$ denotes the eigenvalue of the *light* grand spin $\boldsymbol{g} = \boldsymbol{I} + \boldsymbol{L} + \boldsymbol{S}'$ and $g_3$ its projection. In this definition $\boldsymbol{I}, \boldsymbol{L}$ and $\boldsymbol{S}'$ label the isospin, orbital angular momentum and spin of the light quark inside the heavy meson. The wave–functions $\bar{\Psi}_{dl}$ are most conveniently constructed by introducing the intermediate vector sum $\boldsymbol{K} = \boldsymbol{I} + \boldsymbol{S}'$ possessing the eigenfunctions $\xi_{dl}(k, k_3)$, which represent the product representation of spin and isospin 1/2 objects [9]. Then the eigenfunctions of $\boldsymbol{g}$ are decomposed as

$$\bar{\Psi}_{dl}\left(g, g_3; \hat{\boldsymbol{r}}, k\right) = \sum_{r_3 k_3} C^{gg_3}_{rr_3, kk_3} Y_{rr_3}\left(\hat{\boldsymbol{r}}\right) \xi_{dl}(k, k_3). \tag{3.8}$$

Here $C^{gg_3}_{rr_3, kk_3}$ is the Clebsch–Gordon coefficient associated with the coupling of $r$ and $k$ to $g$. Furthermore $Y_{rr_3}(\hat{\boldsymbol{r}})$ denotes a spherical harmonic function, *i.e.* an eigenfunction of $\boldsymbol{L}^2$ and $L_3$ with eigenvalues $r(r+1)$ and $r_3$, respectively. The spinors are related to those in (2.9) and (2.10) by $\epsilon_{lh}\chi'_h = \chi_l$.

---

[a] Repeated indices are summed over.
[b] In the heavy limit $u^2(r)$ may be considered as a $\delta(r)$–type function or derivatives thereof.

In the rest frame the pseudoscalar and vector parts are finally extracted via

$$P'^{a\dagger} = e^{-i\epsilon t}P^{a\dagger} = \frac{i}{2}\bar{H}^a_{ll} \quad \text{and} \quad Q'^{a\dagger}_i = e^{-i\epsilon t}Q^{a\dagger}_i = -\frac{1}{2}(\sigma_i)_{hl}\bar{H}^a_{lh}. \tag{3.9}$$

Remember that in the heavy limit $Q^a_0 = 0$, cf. eq (2.11). In order to obtain states with good grand spin eigenvalues (as the ansätze (2.9) and (2.10)) one furthermore has to couple the spin $\boldsymbol{S''}$ of the heavy quark according to $\boldsymbol{G} = \boldsymbol{g} + \boldsymbol{S''}$.

It is then straightforward to verify that the relation (3.3), which was obtained by solving the bound state equation in the heavy mass limit, corresponds to the state with the quantum numbers $g = r = k = 0$ while (3.4) is associated to the state[c] with $g = r = 1$ and $k = 0$.

## 3.1. Pentaquark States

In addition to these bound states in the $k = 0$ channel, which carry positive energy eigenvalues, solutions with $G = 1/2$ exist in the $k = 1$ channel possessing, however, negative energy eigenvalues, the so–called pentaquark states [8]. These solutions describe an anti–heavy meson bound to the Skyrmion and correspond to $(qqqq\bar{Q})$ states in the quark model. In the limit $M = M^* \to \infty$ their binding energy is given by

$$\epsilon^{(p)}_B = \frac{1}{2}dF'(0) + \frac{\sqrt{2}c}{gm_V}G''(0) + \frac{\alpha}{2}\omega(0). \tag{3.10}$$

In the heavy limit there are four degenerate negative energy eigenstates with $G = 1/2$. We list the quantum numbers of these states as well as the resulting relations between the radial functions $\Phi, \Psi_1$ and $\Psi_2$.

$$\begin{aligned}
&1. \ g = 0, \ r = 1: \quad \Phi = \Psi_1 = \Psi_2 \quad &&(\text{S} - \text{wave}), \\
&2. \ g = 1, \ r = 1: \quad \Phi = -\Psi_1, \ \Psi_2 = 0 \quad &&(\text{S} - \text{wave}), \\
&3. \ g = 1, \ r = 0: \quad \Phi = -3\Psi_1 = -3\Psi_2/2 \quad &&(\text{P} - \text{wave}), \\
&4. \ g = 1, \ r = 2: \quad \Phi = 0, \ \Psi_1 = -\Psi_2 \quad &&(\text{P} - \text{wave}).
\end{aligned} \tag{3.11}$$

According to the above described scheme these four states may couple to $k = 1$ states because $\boldsymbol{K} = \boldsymbol{g} + (-\boldsymbol{L})$. It should be noted that the relations (3.11) for the heavy limit have to be supplemented by $\Psi_0 = 0$.

Actually there is another way to look at the penta quark states. One may consider these states as the particle conjugated system of heavy mesons being bound to light anti–solitons. This anti–soliton is related to the soliton by [21]

$$F(r) \to -F(r), \quad G(r) \to G(r) \quad \text{and} \quad \omega(r) \to -\omega(r). \tag{3.12}$$

---

[c]Using (3.7)-(3.9) for the $g = r = 1, k = 0$ state yields $P'^\dagger = B\hat{\boldsymbol{r}}(\hat{\boldsymbol{r}} \cdot \boldsymbol{\tau})\chi$ and $Q'^\dagger_i = -iB\hat{\boldsymbol{r}}(\hat{r}_i - i\epsilon_{ijk}\hat{r}_j\tau_k)\chi$ where B is a constant. We note that the overall factor $\hat{\boldsymbol{r}}$ corresponds to the fact that $g = 1$. We couple these fields to $G = 1/2$ by replacing $\chi \to \boldsymbol{\tau}\chi$ and dotting it into the overall $\hat{\boldsymbol{r}}$.This gives $P'^\dagger = B\chi, Q'^\dagger_i = -iB(2\hat{r}_i\hat{\boldsymbol{r}} \cdot \boldsymbol{\tau} - \tau_i)\chi$ which may be compared with (2.10).

For the heavy meson fields the effect of G–conjugation on the radial fields is

$$\Phi(r) \to -\Phi(r), \quad \Psi_0(r) \to -\Psi_0(r) \quad \text{and} \quad \Psi_i(r) \to \Psi_i(r) \tag{3.13}$$

up to an overall sign. It can easily be verified that the Lagrangians (A.1) and (A.3) are invariant under the combined transformations (3.12,3.13) when in addition the sign of the bound state energy ($\epsilon$) is reversed. This, however, just represents the transformation from a bound meson to a bound anti–meson and vice versa. The consideration of the particle conjugation on the penta quarks is useful since it allows one to apply the expression

$$M - \epsilon = \left[3 - 2k^2\right] \left\{ \frac{d}{2} F'(0) - \frac{\sqrt{2}c}{gm_V} G''(0) \right\} + \frac{\alpha}{2} \omega(0) \tag{3.14}$$

found in ref. [9] for the bound state energy of the meson ($\epsilon > 0$) to that of anti–mesons as well. Here $k$ refers to the eigenvalue of $\boldsymbol{K}$. Clearly the application of the particle conjugation prescription to (3.14) (that is, setting $\epsilon \to -\epsilon, F \to -F, G \to G, \omega \to -\omega$) results in (3.10) for $k = 1$. Due to its isoscalar character the $\omega$ field contributes in the same way to the binding energies of both $k = 0$ and $k = 1$ states. Note that (3.13) shows there is a reversal of the sign of $\Phi$ with respect to the $\Psi_i$ when conjugating the heavy limit wave functions obtained from (3.7)-(3.9).

## 4. Boundary conditions

In this section we will examine the compatibility of the heavy quark relations found in the preceding section with the boundary conditions resulting from the bound state equations. This is interesting because the heavy quark relations are determined from the small $r$–behavior without respect to the associated boundary conditions. As the heavy quark relations originate from purely geometrical considerations the boundary conditions may impose additional conditions on the wave-functions related to the dynamics of the system. For simplicity we will omit the light vector mesons $\rho$ and $\omega$ (Their effects will be included in section 6.). We then get the Skyrme model [22] as the one which supports the soliton for the light pseudoscalars:

$$\mathcal{L}_{Sk} = f_\pi^2 \, \text{tr}\,(p_\mu p^\mu) + \frac{m_\pi^2 f_\pi^2}{2} \text{tr}\left[U + U^\dagger - 2\right] + \frac{1}{2g^2} \, \text{tr}\,([p_\mu, p_\nu][p^\mu, p^\nu]), \tag{4.1}$$

where, for simplicity, we have adopted the static limit for the $\rho$ meson, which determines the coefficient of the fourth order stabilizing term. In this model the heavy limit bound state energies are obtained from eqs (3.2) and (3.10) by taking $c = \alpha = 0$.

In order to construct the bound state solutions it is useful to study the differential equations stemming from (A.1) and (A.3) for the boundaries $r \to \infty$ and $r \approx 0$. In the former case the soliton disappears and the differential equations reduce to Klein–Gordan (for $P$) and Proca (for $Q_\mu$) equations in the S– and P wave channels, respectively. This implies that at large $r$ the heavy meson profiles decay like

$$\Phi \sim \exp\left(-\sqrt{M^2 - \epsilon^2}\, r\right), \quad \Psi_\mu \sim \exp\left(-\sqrt{M^{*2} - \epsilon^2}\, r\right). \tag{4.2}$$

For $r \approx 0$ the situation is more complicated since, due to the presence of the soliton, $R_\alpha \approx -2 - (F'(0)r)^2/2$ and the roles of S- and P waves are exchanged. Furthermore one should note that for large but finite $M$ and $M^*$ there is always a vicinity of the origin with $1/r \gg M, M^*$. It is especially illuminating to explore the constraint for $\Psi_1$, the radial component of the heavy vector meson, in this region because it provides access to the relations between $\Phi, \Psi_1$ and $\Psi_2$ without solving a complicated differential equation. For the S wave we obtain for the dominant piece in the vicinity of the origin (again we ignore the time component of the vector field)

$$\begin{aligned} 0 &= 2\left(M^{*2} - \epsilon^2\right)\Psi_1 + \frac{1}{r^2}(R_\alpha + 2)^2\Psi_1 - \frac{2}{r^2}(R_\alpha + 2)\Psi_2 - \frac{4d}{r}\epsilon\sin F\Psi_2 + 2MdF'\Phi + \ldots \\ &\approx 2\left(M^{*2} - \epsilon^2\right)\Psi_1 + F'^2(0)\Psi_2 + 4d\epsilon F'(0)\Psi_2 + 2MdF'(0)\Phi \,. \end{aligned} \quad (4.3)$$

Two observations can be made from this constraint. First we notice that the additional term $F'^2(0)\Psi_2$ may be ignored in the large $M$ limit. Second the remainder is compatible with the relations (3.2, 3.4) for the bound heavy mesons as well as the relations (3.10, 3.11) for the penta quark states. Note that in the latter case $\epsilon \approx -M$. Thus we conclude that in the S wave channel the small $r$ behavior is in agreement with the large $M$ limit. In particular one cannot deduce additional conditions on the wave–functions from the study of the vicinity of the origin. The situation is different in the P wave channel. Here the leading contribution to the constraint for $\Psi_1$ at $r \approx 0$ is found to be

$$\begin{aligned} 0 &= \frac{R_\alpha}{r^2}[R_\alpha\Psi_1 + \Psi_2 + r\Psi_2'] + 2\left(M^{*2} - \epsilon^2\right)\Psi_1 + 2MdF'\Phi + \frac{2d}{r}\epsilon\sin F\Psi_2 + \ldots \\ &\approx \frac{2}{r^2}[2\Psi_1 - \Psi_2 - r\Psi_2'] + \frac{F'^2(0)}{2}[8\Psi_1 - \Psi_2 - r\Psi_2'] \\ &\quad + 2\left(M^{*2} - \epsilon^2\right)\Psi_1 + 2dF'(0)\left(M\Phi - \epsilon\Psi_2\right). \end{aligned} \quad (4.4)$$

Evidently the heavy quark limit results discussed in the previous chapter agree with (4.4) only when the wave–functions behave such that

$$2\Psi_1 - \Psi_2 - r\Psi_2' = \mathcal{O}\left(r^2 \times \Psi_i\right) \quad \text{for} \quad r \approx 0 \quad (4.5)$$

*i.e.* the small $r$ behavior imposes an additional condition which in general may not be compatible with the heavy quark limit. Stated otherwise: The limits $M \to \infty$ and $r \to 0$ do not necessarily commute. The question arises whether this additional constraint can be accommodated by the wave–functions obtained from the "geometrical" considerations studied above in the heavy quark limit. One possible solution to (4.5) is represented by the small $r$ behavior

$$\Phi(r) = \Phi(0) + \mathcal{O}\left(r^2\right), \quad \Psi_1(r) = -\Phi(0) + \mathcal{O}\left(r^2\right) \quad \text{and} \quad \Psi_2(r) = -2\Phi(0) + \mathcal{O}\left(r^2\right). \quad (4.6)$$

Of course, $\Phi(0) \neq 0$ just originates from the above mentioned fact that due to the presence of the soliton the P wave *ansatz* exhibits an S wave behavior at the origin. Obviously the solution (4.6) is compatible with the heavy quark results (3.3) and the $g=1$, $r=0$ state in (3.11). The relation (4.5) can also be satisfied by the $g=1$, $r=2$ state in (3.11) when

$$\Psi_1(r) = \frac{\text{const.}}{r^3} + \ldots \quad \text{and} \quad \Psi_2(r) = -\frac{\text{const.}}{r^3} + \ldots , \qquad (4.7)$$

while $\Phi$ vanishes. On the other hand this solution is highly singular at the origin and should be discarded because it is apparently not normalizable[a]. We therefore conclude that the bound penta quark state with $g=1$, $r=2$ and coupled[b] to $G=1/2$ is forbidden by the dynamics of the system, although the geometry of the heavy quark limit indicates that this state may exist. However, as we are interested in finite (but large) masses, the solutions to the bound state equations always have to satisfy the appropriate boundary conditions at $r=0$.

Table 5.1: The bound state energies of the states explored in the previous chapters. Displayed are the lowest radial excitations in each channel only. The data in parentheses in the $k=0$ channel refer to the results obtained using the approximate bound state equation (A.9).

| | | $k=0$ $\epsilon_B$ (MeV) | | $k=1$ $\epsilon_B^{(p)}$ (MeV) | | |
|---|---|---|---|---|---|---|
| heavy limit | | 1016 | | 339 | | |
| $M$(GeV) | $M^*$(GeV) | $g=0, r=0$ | $g=1, r=1$ | $g=0, r=1$ | $g=1, r=1$ | $g=1, r=0$ |
| 50.0 | 50.0 | 869 (866) | 769 (782) | 169 | 231 | 260 |
| 40.0 | 40.0 | 853 (850) | 743 (758) | 153 | 220 | 252 |
| 30.0 | 30.0 | 831 (828) | 706 (725) | 130 | 206 | 241 |
| 20.0 | 20.0 | 796 (790) | 646 (674) | 96 | 183 | 222 |
| 10.0 | 10.0 | 721 (709) | 519 (570) | 35 | 136 | 182 |
| 5.279 | 5.325 | 595 (608) | 338 (457) | — | 71 | 118 |
| 1.865 | 2.007 | 314 (353) | 29 (239) | — | — | — |

## 5. Numerical Results for Finite Heavy Meson Masses

In this section we will investigate the behavior of the bound heavy mesons when going from the large $M$ limit to the realistic values $M=5.279$GeV (1.865GeV) and $M^*=5.325$GeV (2.007GeV) in the $B(D)$ meson system. It will be of special interest whether or not the bound penta quark states discussed in the previous sections persist in the realistic cases. Again we will restrict ourselves to the Skyrme model for the light sector. In this case we arbitrarily choose the Skyrme coupling constant so that the P wave binding energy (3.2) in the heavy limit coincides with the value obtained in the model including light vector mesons as well (cf. Sect. 6). This procedure yields $g=6.45$. We will consider the pion decay constant, $f_\pi$ to have its experimental value. Kinematic corrections due to the finite nucleon mass will not yet be included.

---

[a]To make this argument more precise we would have to furnish a suitable metric.
[b]Further investigation is required for the case when the $g=1, r=2$ state is coupled to $G=3/2$.

In the case $M \neq M^*$ we define the binding energy with respect to $M$. Our main numerical results are summarized in table 5.1. We observe that the heavy limit results for the binding energies are only slowly approached as the masses increase. Even for $M = 50$ GeV the infinite M approximation is not good in this respect. For realistic masses we find that the $g = 0, r = 1$ penta quark state becomes unbound in the B–meson system while the two other allowed penta quark states remain bound. Turning to the D–meson system all these penta states get shifted into the continuum.

In ref. [9] an approximate bound state equation was derived by substituting the relations (3.4) and (3.3) into the associated equations of motion for $\Phi$. In the appendix this approach is repeated for the case when the light vector mesons are also present, yielding the approximate bound state equation (A.9). Clearly this approximation provides excellent agreement with the exact result in the case of the P wave. Even for masses as small as in the D–meson system it represents a useful guideline. For the S wave the solution to (A.9) can be considered as an estimate for the upper bound for the binding energy but for masses as small as in the D–meson sector this approximation fails.

Next let us examine the radial wave–functions for the states under study. We observe from figure 5.1 that at $M = 50$ GeV, in contrast to the binding energy predictions, the heavy limit relations (3.3) and (3.4) for the wave–functions are remarkably well satisfied. Also the time component of the vector meson field is significantly reduced as compared to the other radial functions. Figure 5.2 shows that the heavy limit relation for the $g = 1, r = 1$ penta quark state is also well reproduced. This is in contrast to the situation for the $g = 0, r = 1$ state ( see the first relation in eq (3.11)). There, at the maximum $\Phi$ is almost twice as large as $\Psi_2$. Furthermore $\Psi$ develops a node at larger distances while neither $\Psi_1$ nor $\Psi_2$ do. As can be seen from figure 5.3 the P wave penta quark state with the quantum numbers $g = 1$, $r = 0$ satisfies the associated relation in eq (3.11) only in the vicinity of the origin. Turning now to realistic values for the masses of the heavy mesons figure 5.4 shows that this pattern persists and that the bound mesons reasonably well satisfy the heavy limit relations (3.3) and (3.4), although at least for the S wave the time component of the vector meson becomes quite pronounced. Again the heavy limit relations for the bound penta quark states are only approximately satisfied, *cf.* figure 5.5.

We should also mention that for realistic masses the relation (2.11) is satisfied at the 5% level for the S wave. For the P wave the discrepancy is larger; however, in the physically relevant region both sides of that equation turn out to be about three orders of magnitude smaller than the amplitudes of the other radial functions. Hence we conclude that (2.11) represents a justifiable approximation.

In the above computations we have fixed the Skyrme coupling constant by requiring the same heavy limit binding energy $\epsilon_B = 1016$ MeV as we get for the vector mesons (see Sect. 6). This yields $g = 6.45$. Then the binding energies in the realistic cases are predicted to be too small. Alternatively one may demand the experimental value (780 MeV) for the binding of $\Lambda_b$. Then we obtain a value as large as $g = 9.50$. For this value the binding energy of the $\Lambda_c$ is found to be 370 MeV which underestimates the experimental value (630 MeV) by almost a factor of two. In any event the pseudoscalar soliton with $g = 9.50$ has to be abandoned because it describes the properties of the light baryons very badly. For example, the mass difference between the $\Delta$ resonance and the nucleon is obtained to be 2.67 GeV!

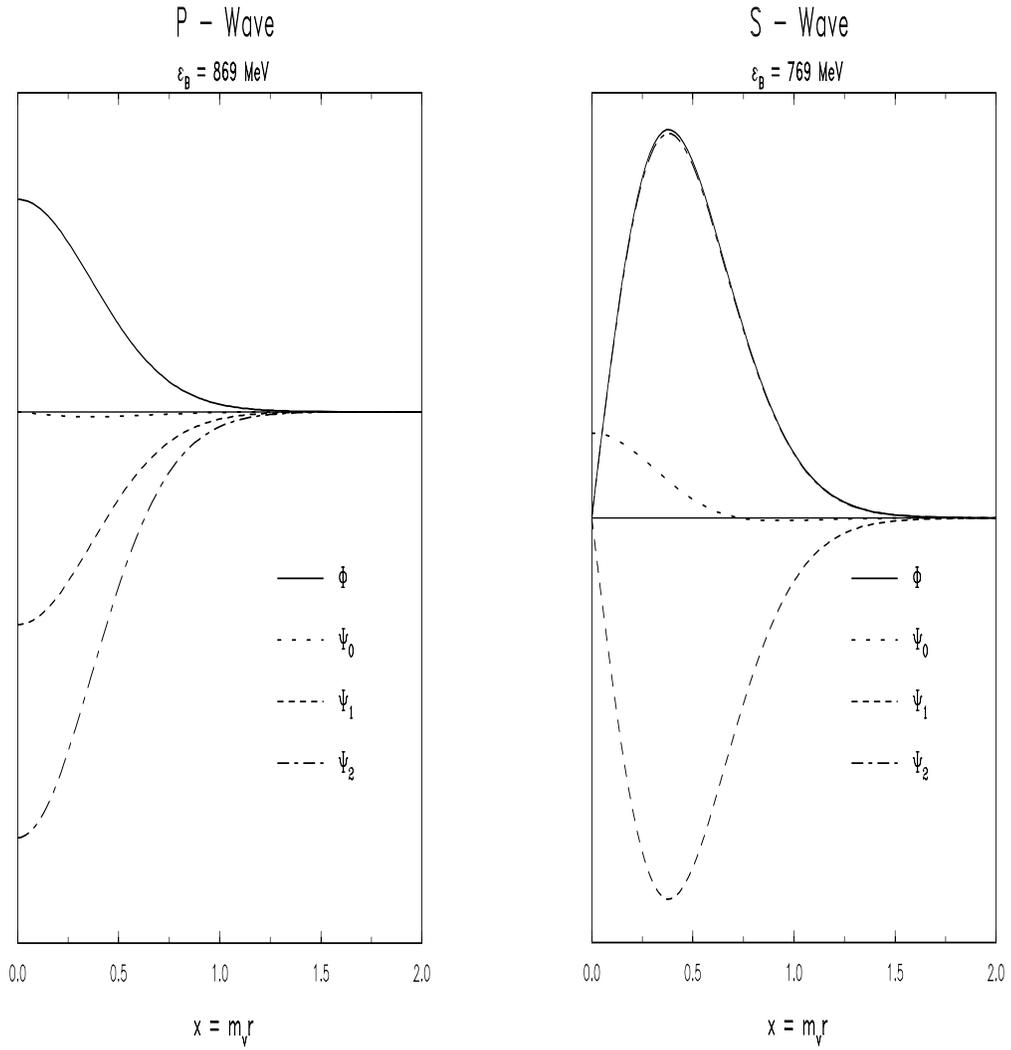

Figure 5.1: The profiles for the lowest radial excitations of the bound heavy P (left panel) and S (right panel) wave mesons in the case $M = M^* = 50 \text{GeV}$. The normalization of the wave–functions is chosen arbitrarily.

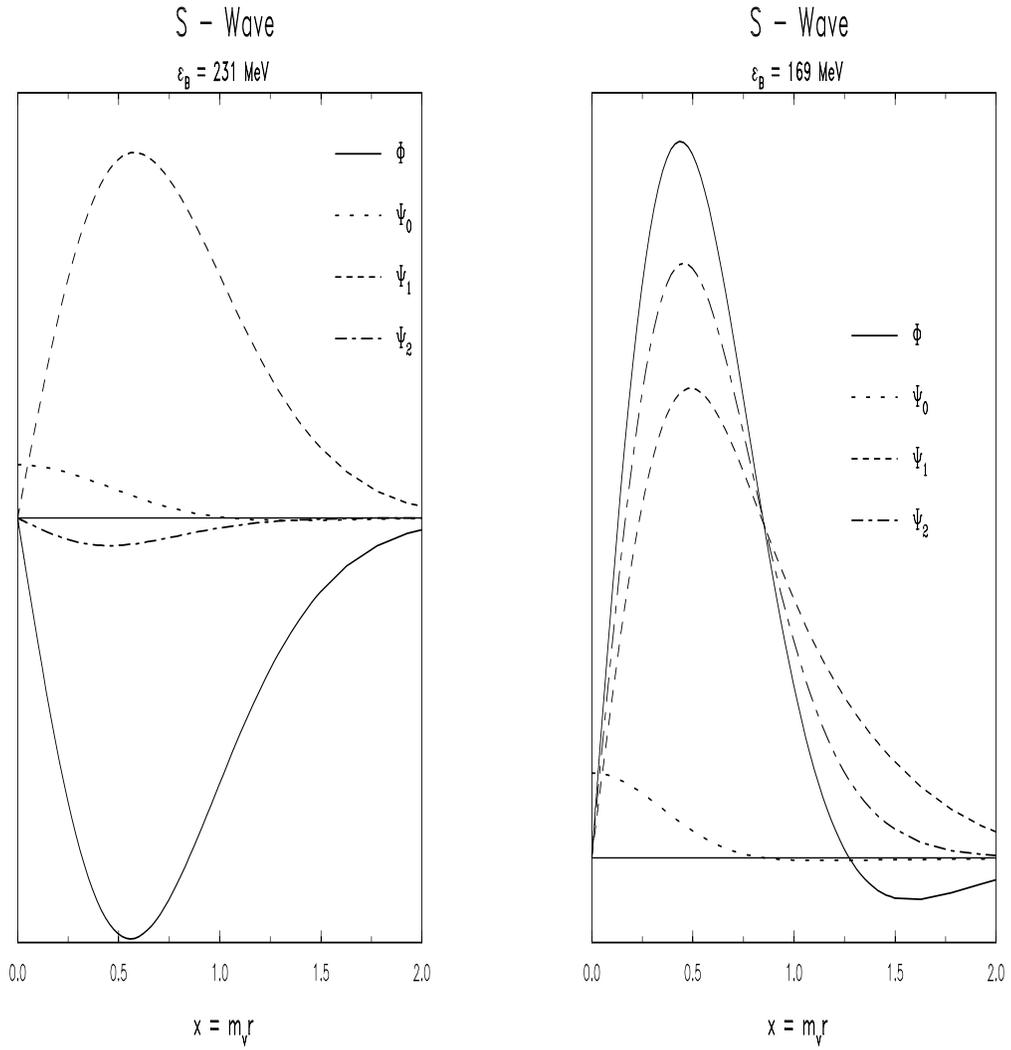

Figure 5.2: The profiles for the lowest bound heavy S wave anti–mesons in the case $M = M^* = 50\text{GeV}$. Left panel: $g = 1, r = 1$, right panel: $g = 0, r = 1$. The normalization is chosen arbitrarily.

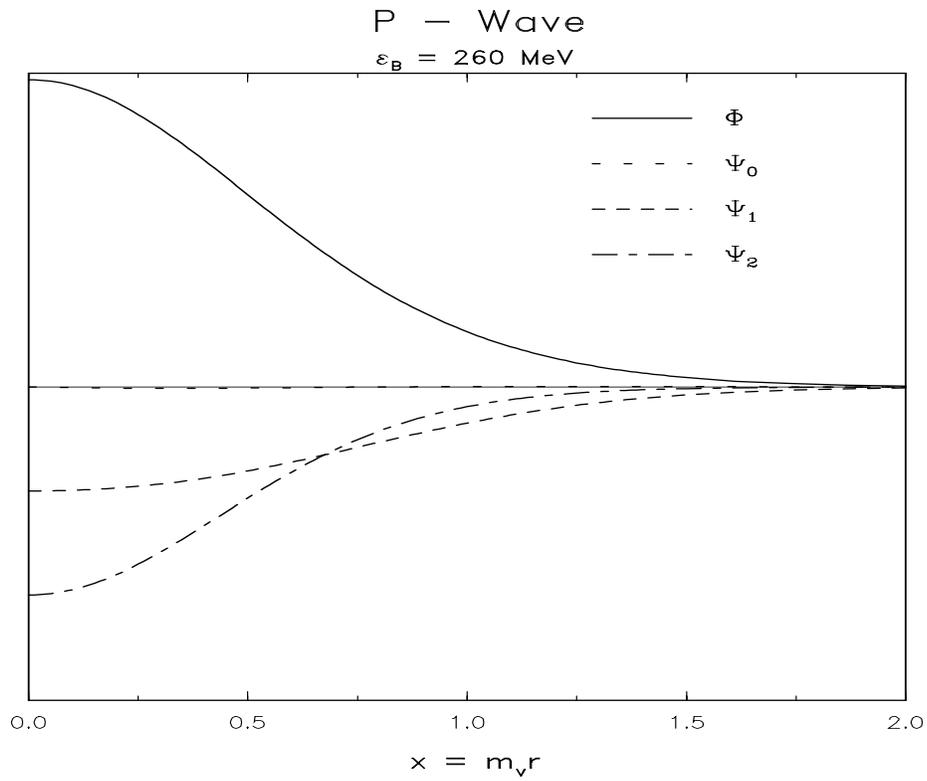

Figure 5.3: The profiles for the lowest radial excitation of the bound heavy P wave anti–meson in the case $M = M^* = 50\text{GeV}$. The normalization is chosen arbitrarily.

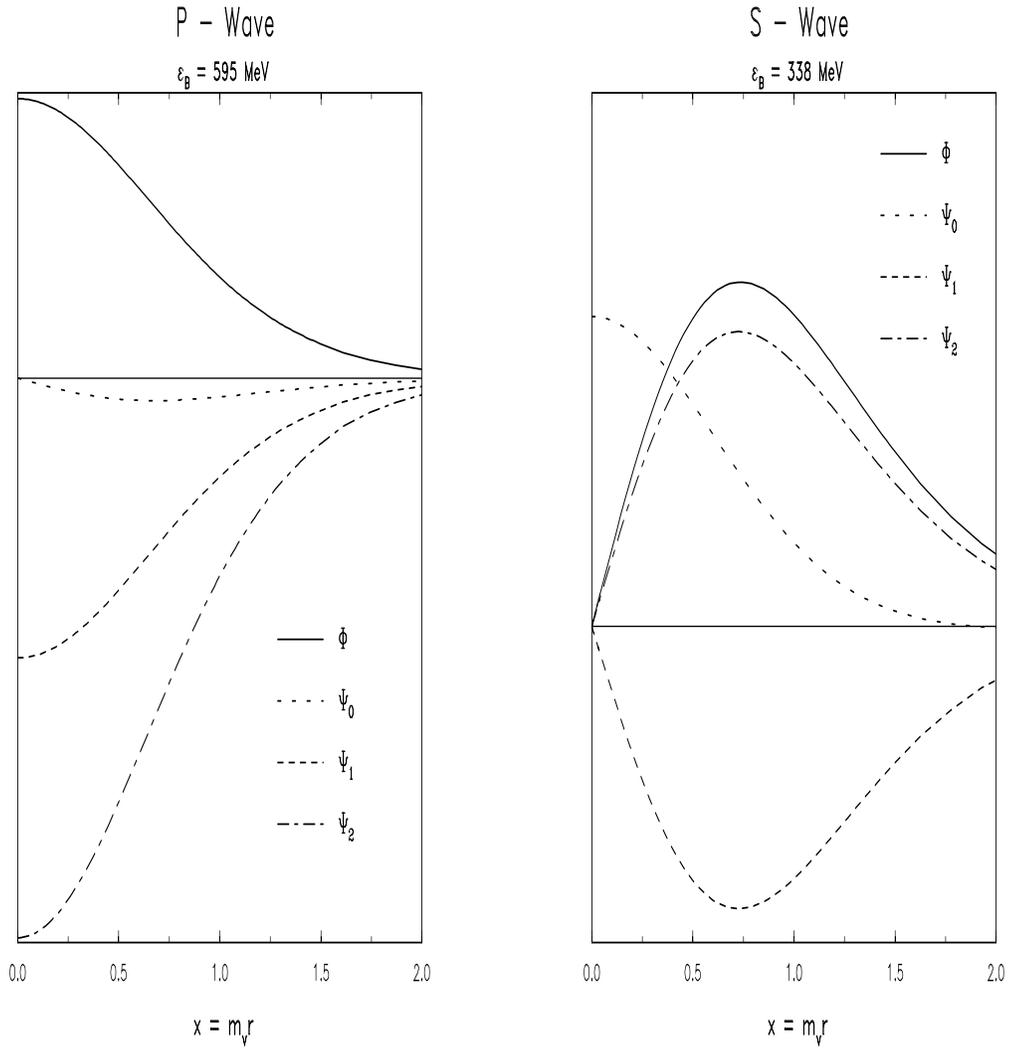

Figure 5.4: The profiles for the lowest radial excitations of the bound heavy P (left panel) and S (right panel) wave mesons in the case $M = 5.279\text{GeV}$ and $M^* = 5.325\text{GeV}$. The normalization of the wave–functions is chosen arbitrarily.

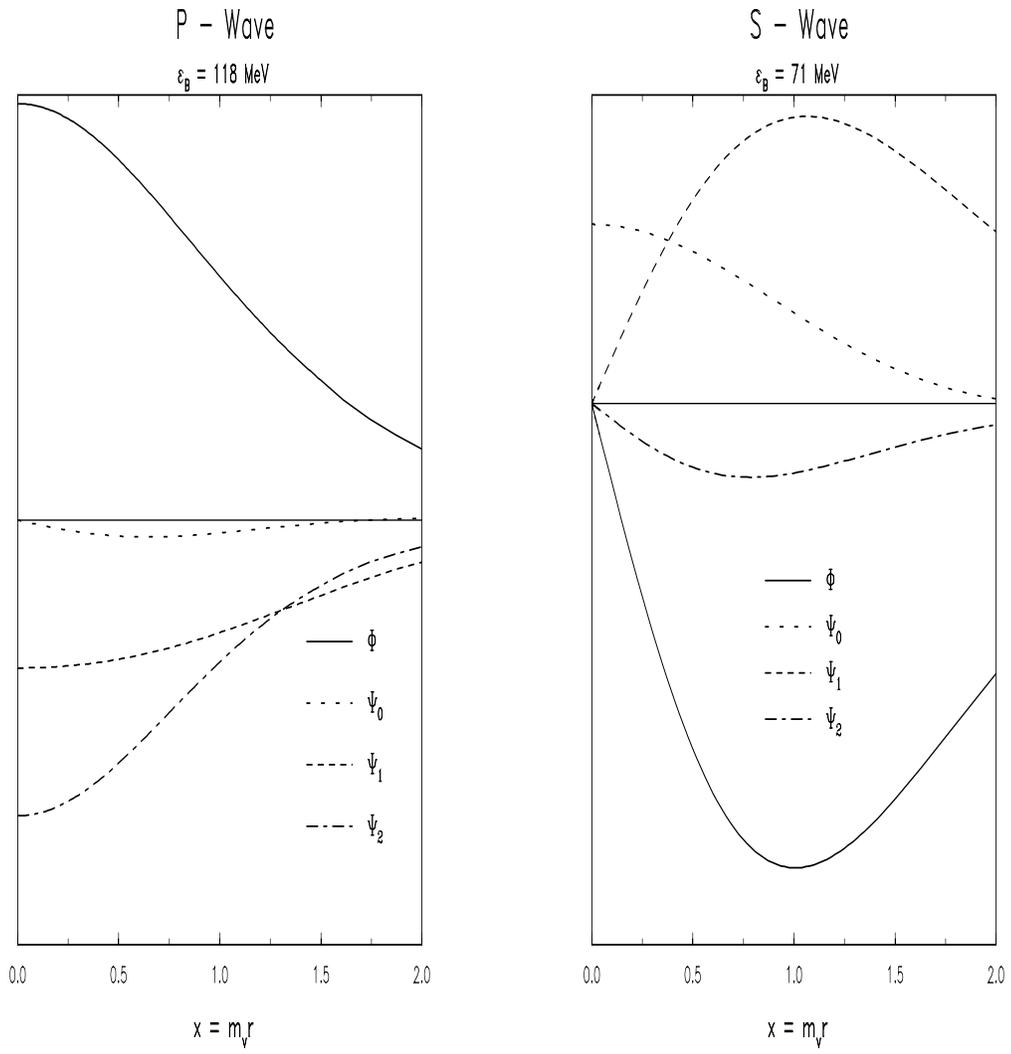

Figure 5.5: Same as figure 5.4 for the bound heavy anti–mesons.

As a reminder we note that for $g = 6.45$ this difference comes out to be 905 MeV which also is far too large. Fixing $g$ from the experimental value (293 MeV) for this mass difference yields $g = 4.30$. The associated results for the binding energies of the $\Lambda_b$ (443 MeV) and $\Lambda_c$ (244 MeV) baryons are far too small.

## 6. Including Light Vector Mesons

We have just seen that it is very difficult to achieve a consistent picture of both the light and heavy sectors in the model where the soliton is supported by the pseudoscalars only. This motivates the inclusion of light vector mesons, especially since we then have one more undetermined parameter in the heavy sector ($\alpha$), which can be employed to fix the binding energy of $\Lambda_b$ while keeping the parameters in the light sector untouched. However, as soon as we include the vector mesons we encounter the problem that for $\alpha \neq 0$ the coefficient $f_1$ in the effective bound state Lagrangian(A.5) may develop a node, i.e.

$$M^2 - \left(\epsilon - \frac{\alpha}{2}\omega\right)^2 \tag{6.1}$$

may change its sign in between $r = 0$ and $r \to \infty$. This in turn causes the Lagrangian (A.6) to become singular. Since we have $\omega(r) \geq 0$ this problem occurs for the bound heavy mesons ($\epsilon > 0$) when $\alpha < 0$ and for the bound anti-heavy mesons ($\epsilon < 0$) when $\alpha > 0$. In the former case the binding energy is large enough so that the singularity fortunately does not appear in the considered range $\alpha \geq -1$. In the latter case, however, numerical instabilities occur for $\alpha \approx 1$. In order to avoid this numerical problem we will omit the discussion of penta quark states for sets of parameters which lead to nodes of (6.1).

In table 6.1 the dependence of the binding energies in the S and P wave channels on $\alpha$ are displayed. The parameters for the light sector of the model

$$g = 5.57, \qquad m_V = 773\,\text{MeV}$$
$$\gamma_1 = 0.3, \quad \gamma_2 = 1.8, \quad \gamma_3 = 1.2, \tag{6.2}$$

have been taken from ref. [11]. These were determined to provide a best fit to the mass differences of the low–lying $\frac{1}{2}^+$ and $\frac{3}{2}^+$ light baryons, allowing variation only for those parameters which could not be determined from the light meson sector [17]. Furthermore various static properties have been reasonably well described using these parameters [12].

Again we observe that in the P-wave case the approximation (A.9) is very good while in the S wave channel the corresponding solution overestimates the exact binding energy. In this respect the behavior of the model is similar to the one without light vector mesons.

We remark that for $\alpha \approx 0$ numerically stable penta quark solutions in both S and P wave channels are observed. However, these carry binding energies no larger than about 50MeV and hence may well disappear upon modification of the model. The vanishing binding of the penta–quarks for $\alpha \approx 0$ has already been conjectured in ref. [9] from the fact that $\frac{d}{2}F'(0) + \frac{\sqrt{2}c}{gm_V}G''(0)$ in eq (3.10) deviates only slightly from zero for the parameters (6.2).

Of course, the number of bound radial excitations also depends on $\alpha$. Our predictions for the binding energies of these radial excitations are shown in table 6.2.

Table 6.1: The dependence of the binding energy on the undetermined parameter $\alpha$. Data are given in MeV. In the heavy sector we have adopted the parameters (2.8) for the B–meson. The data in parentheses refer to the results obtained using the approximate bound state equation (A.9).

| $\alpha$ | P wave | S wave |
|---|---|---|
| -1.00 | 1035 (1036) | 847 (900) |
| -0.75 | 972 (976) | 791 (843) |
| -0.50 | 910 (916) | 735 (785) |
| -0.25 | 848 (857) | 679 (729) |
| 0.00 | 786 (797) | 624 (672) |
| 0.25 | 725 (738) | 569 (617) |
| 0.50 | 664 (679) | 515 (562) |
| 0.75 | 604 (621) | 462 (507) |
| 1.00 | 544 (563) | 410 (454) |

Table 6.2: The binding energies (in MeV) of the ground state and radial excitations as functions of $\alpha$ for the S– and P wave channels. Again the parameters for the B–meson (2.8) have been used to illustrate this dependence.

| $\alpha = -0.5$ | | $\alpha = 0.0$ | | $\alpha = 0.5$ | |
|---|---|---|---|---|---|
| P–wave | S–wave | P–wave | S–wave | P–wave | S–wave |
| 910 | 735 | 786 | 624 | 664 | 515 |
| 568 | 428 | 472 | 344 | 379 | 264 |
| 299 | 196 | 229 | 139 | 165 | 88 |
| 109 | 45 | 66 | 18 | 30 | — |
| 34 | 27 | — | — | — | — |

Let us adjust $\alpha$ so that the binding energy of the P wave baryon $\Lambda_b$ is reproduced, *i.e.* $\epsilon_{\Lambda_b} = 780$ MeV. From table 6.2 we observe that this yields $\alpha \approx 0$; the exact value reads $\alpha = 0.03$. As a prediction we then find $\epsilon_{\Lambda_c} = 536$ MeV which is somewhat smaller than the experimental value 630 MeV but certainly an improvement compared to the pseudoscalar soliton. The resulting wave–functions are displayed in figure 6.1. In the bottom sector we then find a penta quark state which is bound by 18 MeV. No bound penta quark state is observed in the charm sector.

In the S wave channel we observe bound states with binding energies 301 MeV and 617 MeV in the charm and bottom sectors, respectively. Again a penta quark state is found only in the bottom sector. This state is bound by 19 MeV.

Since the observed penta quark states are only very weakly bound it is suggestive that these states might disappear once kinematical corrections due to the finite mass of the soliton are incorporated. These corrections will be the subject of the next section.

## 7. *Estimate of Kinematical Corrections*

Following ref. [9] we attempt to estimate the corrections related to the finiteness of the soliton mass by substituting for $M$ and $M^*$ the reduced mass ($\mu$) of the soliton and the heavy meson under consideration into the bound state equations associated with (A.1) and (A.3). For the parameters (6.2) the mass of the soliton turns out to be 1631 MeV[a]. This leads to the reduced masses

$$\begin{aligned} \mu_b &= 1246\,\text{MeV}, \quad \mu_b^* = 1248\,\text{MeV}, \\ \mu_c &= 870\,\text{MeV}, \quad \mu_c^* = 900\,\text{MeV}. \end{aligned} \qquad (7.1)$$

Clearly this estimate of the kinematical corrections provides a major change of the parameters.

We first note that within the model of pseudoscalars only, the resulting binding energies for the P wave states are 326 MeV and 211 MeV in the B–meson and D–meson sectors, respectively. Here we have again used $g = 6.45$. These numbers are significantly smaller than the experimental data. Adopting $g = 4.30$, which gives the correct nucleon–$\Delta$ mass difference, leads to even smaller binding energies. Moreover, in the S wave channel we do not observe any bound states in either the B–meson or in the D–meson sector.

When vector mesons are included we can again make use of the *a priori* undetermined parameter, which we fix by reproducing the binding energy of $\Lambda_b$ for (7.1). This yields $\alpha = -0.97$. In the S wave channel we then observe a bound state with $\epsilon_B = 420$ MeV. More importantly, the binding energy of the charmed baryon is obtained to be 638 MeV, in remarkably good agreement with the experimental value. Clearly, fitting just one parameter in the vector meson model provides a nice overall agreement with the experimental spectrum of the heavy baryons. The resulting wave–functions are displayed in figure 7.1. Although the masses assumed for the heavy mesons (7.1) are quite small the heavy limit relations (3.3) are still approximately satisfied. An S wave state carrying a binding energy of 256 MeV is also obtained in the D–meson sector. This result suggests identifying this state with the recently discovered $\Lambda_c'$, which is supposed to be bound by approximately 320 MeV. These results are summarized in table 7.1. We also remark that the number of radial excitations is quite limited in this case. For example, we observe the first excited P wave state in the B–system just 184 MeV below the threshold. We will not further pursue these radial excitations since (as indicated after (6.1)) states with such low binding energies are troublesome numerically.

---

[a]Quantum corrections [15] will reduce this value so that a reasonable description of the nucleon mass is obtained. These corrections are formally subleading in $1/N_C$.

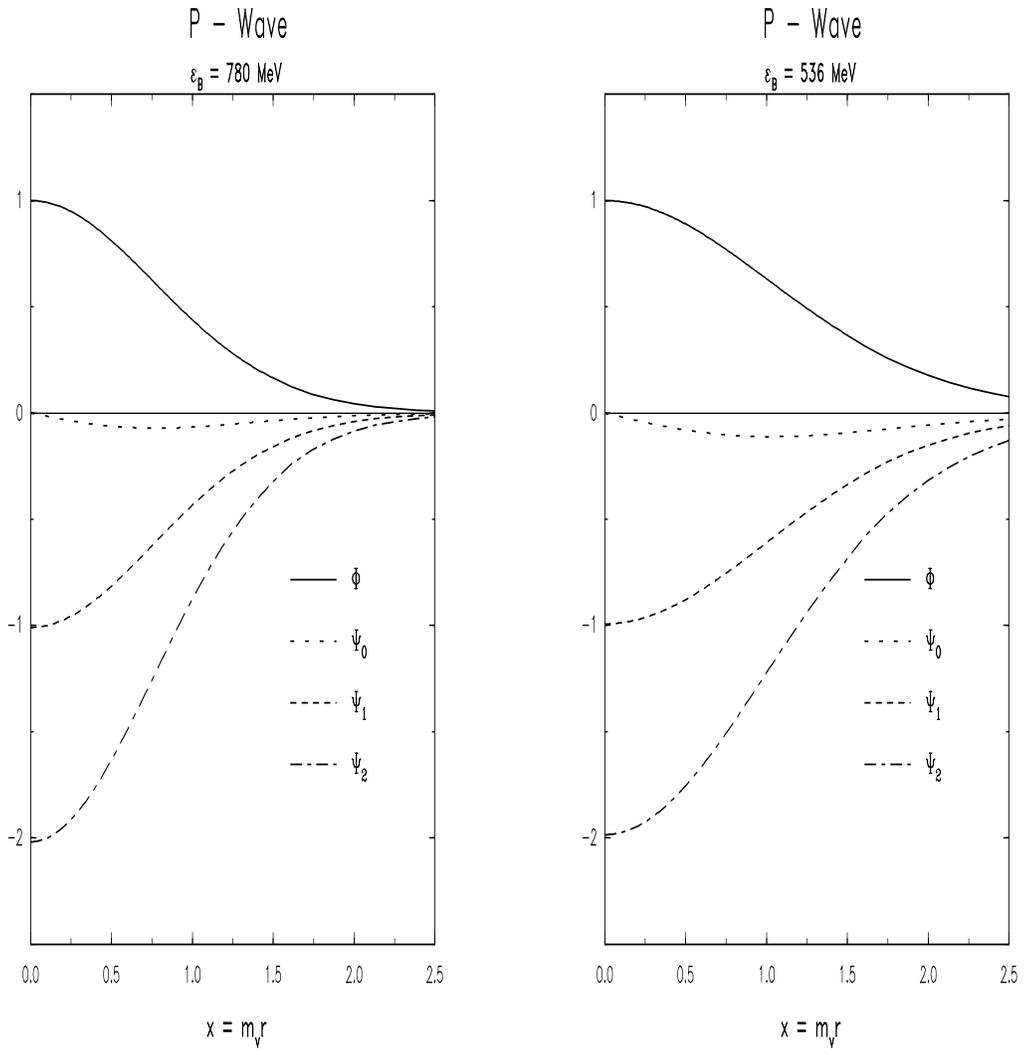

Figure 6.1: The wave–functions for the bound mesons in the bottom (left) and charm (right) sectors.

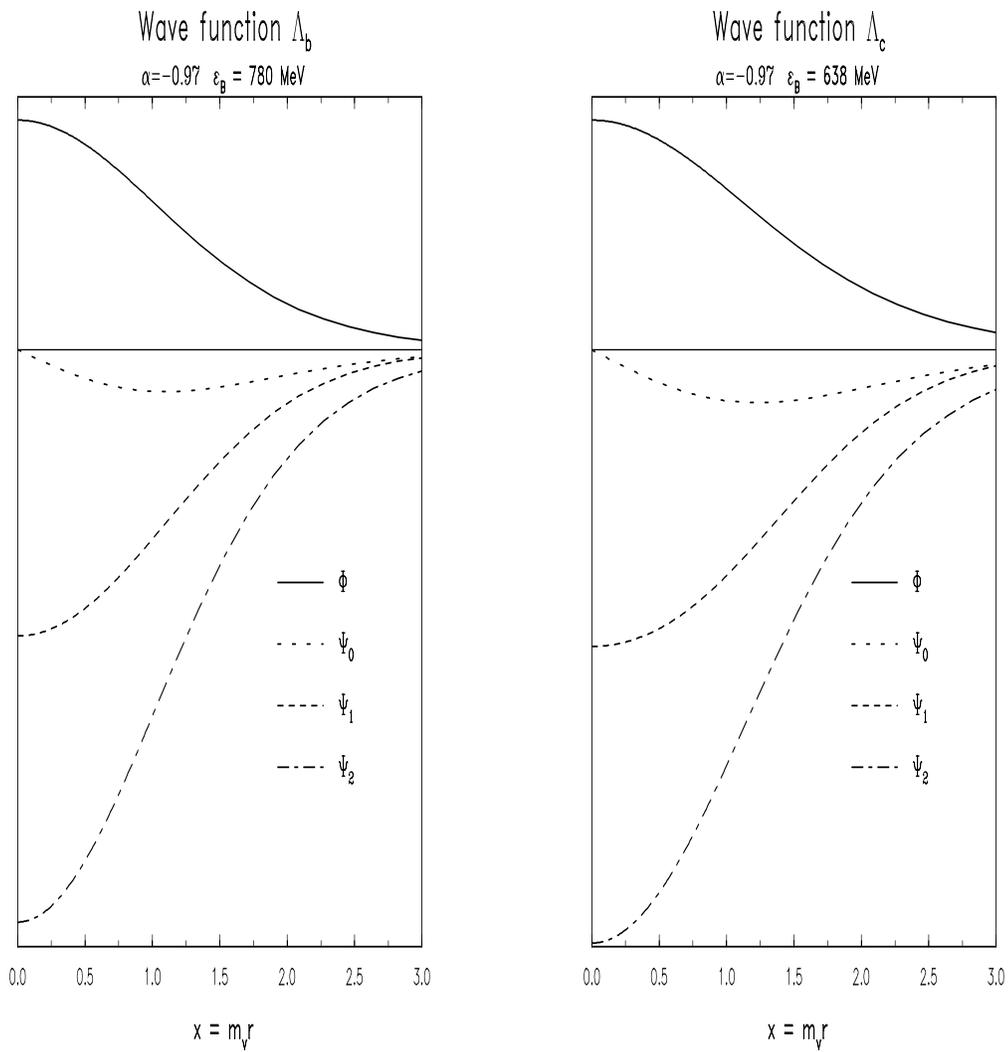

Figure 7.1: The wave–functions for the bound mesons in the bottom (left) and charm (right) sectors in the reduced mass approximation for the kinematical corrections.

Table 7.1: The bound state energies with kinematic corrections taken into account. Displayed are the lowest radial excitations in each channel only, as well as the known experimental data.

|         |              | $c$ baryons | $b$ baryons |
|---------|--------------|-------------|-------------|
| P Wave: | $\epsilon_B$, MeV | 638 | 780 (fit) |
|         | Expt., MeV   | 630 | 780 |
| S Wave: | $\epsilon_B$, MeV | 256 | 420 |
|         | Expt., MeV   | 320 | ? |

As a further consequence of the small value for $\alpha$ no bound penta quark states are observed.

Finally it is worthwhile to note that employing the above described procedure for the kaons, *i.e.* substituting $\mu_K = 380$ MeV and $\mu_K^* = 572$ MeV while keeping $\alpha = -0.97$, leads to a binding energy of 233 MeV for the P wave. Although the heavy quark symmetry is certainly not valid for the kaons this result compares favorably with the isospin weighted average binding energies of the $\Lambda$ and $\Sigma$ hyperons (261 MeV). Without the kinematical corrections we would find a very loosely bound P wave kaon. The reason is that we then need to choose a value for $\alpha$ which is close to zero.

## 8. Summary

In this paper we have considered baryons containing one heavy quark or anti-quark as respectively bound states of heavy mesons or anti-mesons in the background of a soliton, which is constructed from light pseudoscalar and vector mesons. To describe the coupling between the heavy and light mesons we have employed the simplest generalization of the heavy spin symmetric effective interaction to finite masses. No further approximations associated with the heavy mass limit have been made. We have restricted our investigations to the physically interesting cases of S– and P wave heavy mesons. Special emphasis has been put on studying the behavior of the binding energies and the bound state wave–functions when going from the heavy mass limit to physically realistic masses. We have observed that the heavy limit result for the binding energy is actually approached very slowly when increasing the heavy meson mass. For the empirical meson masses the heavy limit result overestimates the exact solution to the bound state equation by a factor of two (three) in the B(D) meson system. Furthermore the degeneracy between the binding energies in the S– and P wave channels is removed. In the case of the antimeson fields (related to the penta quark baryons $q^4\bar{Q}$) we have observed the interesting feature that a state, which in the geometrical coupling scheme of the heavy symmetry theory is predicted to be bound, is prohibited by the dynamics. Technically this is caused by the non–commutativity of the limits $M \to \infty$ and $r \to 0$. Hence this state does not exist even for large but finite masses.

Turning to realistic meson masses, the binding energy of the allowed S– and P wave penta quark states decreases leaving these states unbound in the D meson system.

Although the binding energy deviates significantly from the heavy limit result we noted that in the course of going to realistic masses the P wave radial functions reasonably well satisfy the heavy limit relations. As a consequence the approximative bound state equation obtained by substituting these relations for the heavy vector meson fields in the differential equation for the heavy pseudoscalar field provides a binding energy close to the exact result. This approximation may be a useful simplification in other more complicated models [23].

We found that it was not possible to obtain a reasonable description of the properties of both the light and heavy baryons when the Skyrme model is employed for the light soliton sector[a]. This made mandatory the incorporation of light vector mesons. Although that brings into the game one more parameter, describing the coupling of the isoscalar $\omega$ meson to the heavy meson fields, we observed the satisfying feature that adjusting this parameter to the binding energy of the lowest baryon containing a heavy bottom quark leads to a reasonable agreement with the data available for other baryons with a heavy quark. This picture holds when kinematical corrections are approximated by substituting the reduced masses in the bound state equations. In that case not only the penta quarks in the S wave channel but also those in the P wave channel become unbound. Furthermore the number of radially excited bound states is reduced. In any event, the kinematical corrections require more thorough investigation.

In this paper we neglected the relatively small splitting between heavy baryons of isospin zero and one. The baryons constructed so far carry neither good spin nor isospin quantum numbers. The construction of such states will make it possible to computate this splitting. In order to generate these states collective coordinates $A(t)$ describing the isospace orientation have to be introduced. Specifying the spinor in Fourier space

$$\chi(\epsilon) = (\chi_1(\epsilon), \chi_2(\epsilon)) \tag{8.1}$$

the coupling between the collective rotations and the bound state assumes the form

$$\int dt L_{\boldsymbol{\Omega},\chi} = -\frac{1}{2} \int \frac{d\epsilon}{2\pi} c_h(\epsilon) \left( \sum_{i,j=1}^{2} \chi_i^*(\epsilon) \boldsymbol{\tau}_{ij} \chi_j(\epsilon) \right) \cdot \boldsymbol{\Omega} \tag{8.2}$$

with $\boldsymbol{\Omega} = -i\mathrm{tr}\left(\boldsymbol{\tau} A \dot{A}\right)$ being the angular velocity of the iso–rotations. The coupling coefficient $c_h(\epsilon)$ is bilinear in the radial wave–functions $\Phi$ and $\Psi_\mu$. The subscript $h$ has been attached to denote the heavy flavor under consideration. In order to compute $c_h(\epsilon)$ the wave–functions should be normalized to carry a unit heavy flavor quantum number. Finally the hyperfine splitting will be given by

$$M_{\Sigma_h} - M_{\Lambda_h} = \frac{1 - c_h(\epsilon_B)}{\alpha^2} \tag{8.3}$$

where $\alpha^2$ refers to the moment of inertia corresponding to the isorotation $A$. Upon canonical quantization of $\chi_i(\epsilon)$ the bound state (with energy $\epsilon_B$) is projected out from the Fock space of meson fluctuations. It will be very interesting to pursue this path although in the model including light vector mesons further complications arise because field components, which vanish classically, get induced by this collective rotation [18]. It may also be interesting to examine the effect of translational collective coordinates on the kinematical corrections associated with finite light baryon mass.

---

[a]Parameters should be fit from the meson sector as far as possible.

## *Acknowledgements*

One of us (HW) acknowledges support by a Habilitanden–scholarship of the Deutsche Forschungsgemeinschaft (DFG). This work was supported in part by the US DOE contract number DE-FG-02-85ER 40231 and by the National Science Foundation under Grant PHY-9208386.

## *Appendix A: The Lagrangian for the S– and P–wave heavy mesons*

In this appendix we present the Lagrangian density for the *ansätze* (2.9) and (2.10) of the bound heavy mesons. Due to parity invariance these two channels decouple. We also add some analytical results for the limit of large meson masses.

Substituting (2.10) in (2.5) gives for the S wave channel

$$
\begin{aligned}
\mathcal{L}_H^{(S)} &= \Phi'^2 + \left[ M^2 - \left( \epsilon - \frac{\alpha}{2}\omega \right)^2 + \frac{R_\alpha^2}{2r^2} \right] \Phi^2 + M^{*2} \left[ \Psi_1^2 + 2\Psi_2^2 - \Psi_0^2 \right] \\
&+ \frac{2}{r^2} \left[ r\Psi_2' + \Psi_2 - \frac{R_\alpha + 2}{2} \Psi_1 \right]^2 + \frac{R_\alpha^2}{r^2} \Psi_2^2 - \left[ \Psi_0' - \left( \epsilon - \frac{\alpha}{2}\omega \right) \Psi_1 \right]^2 \\
&- 2 \left[ \left( \epsilon - \frac{\alpha}{2}\omega \right) \Psi_2 - \frac{R_\alpha + 2}{2r} \Psi_0 \right]^2 + 2Md \left[ F'\Psi_1 + \frac{2}{r}\sin F \Psi_2 \right] \Phi \\
&+ 2d \Bigg\{ F' \left[ \frac{1}{r}(1+\cos F) \Psi_0 \Psi_2 - \left( \epsilon - \frac{\alpha}{2}\omega \right) \Psi_2^2 \right] \\
&\quad + \frac{2}{r}\sin F \left[ \Psi_2 \Psi_0' - \left( \epsilon - \frac{\alpha}{2}\omega \right) \Psi_1 \Psi_2 + \frac{R_\alpha + 2}{2r} \Psi_0 \Psi_1 \right] \Bigg\} \\
&+ \frac{4\sqrt{2}cM}{gm_V} \left[ \omega' \Psi_0 \Psi_1 + \frac{2G'}{r} \Psi_1 \Psi_2 + \frac{G}{r^2}(G+2)\Psi_2^2 \right] \\
&+ \frac{4\sqrt{2}c}{gm_V} \Bigg\{ \frac{\omega'}{r} R_\alpha \Phi \Psi_2 + \frac{G}{r^2}(G+2) \Psi_0 \Phi' + \frac{G'}{r} R_\alpha \Phi \Psi_0 \\
&\quad + \left( \epsilon - \frac{\alpha}{2}\omega \right) \left[ \frac{2G'}{r} \Psi_2 + \frac{G}{r^2}(G+2)\Psi_1 \right] \Phi \Bigg\}.
\end{aligned} \quad \text{(A.1)}
$$

Here a prime indicates a derivative with respect to the radial coordinate $r$. Furthermore the abbreviation $R_\alpha = \cos F - 1 + \alpha(1 + G - \cos F)$ has been introduced. It should be noted that we have omitted the overall factor $\chi^\dagger \chi$.

In the limit $M = M^* \to \infty$ the leading contribution is obtained to be

$$\begin{aligned}\mathcal{L}_H^{(S)} \to & \left[M^2 - \left(\epsilon - \frac{\alpha}{2}\omega\right)^2\right]\left[\Phi^2 + \Psi_1^2 + 2\Psi_2^2\right] + 2Md\left[F'\Psi_1 + \frac{2}{r}\sin F \Psi_2\right]\Phi \\ & -2d\left(\epsilon - \frac{\alpha}{2}\omega\right)\left[F'\Psi_2 + \frac{2}{r}\sin F \Psi_1\right]\Psi_2 + \frac{4\sqrt{2}cM}{gm_V}\left[\frac{2G'}{r}\Psi_1 + \frac{G}{r^2}(G+2)\Psi_2\right]\Psi_2 \\ & + \frac{4\sqrt{2}c}{gm_V}\left(\epsilon - \frac{\alpha}{2}\omega\right)\left[\frac{2G'}{r}\Psi_2 + \frac{G}{r^2}(G+2)\Psi_1\right]\Phi. \end{aligned} \quad (A.2)$$

Note that the energy eigenvalue $\epsilon$ is of the order $M$ hence the first term in (A.2) is also of this order. We have additionally used the fact that $\Psi_0$ may be omitted in the heavy limit, *cf.* eq (2.11). Performing the expansion (3.1) leads to the binding energies (3.2) and (3.10) together with the corresponding relations for the wave–functions (3.4) and (3.11) for bound S wave mesons and anti–mesons, respectively.

For the P wave channel one obtains upon substitution of the *ansatz* (2.9)

$$\begin{aligned}\mathcal{L}_H^{(P)} = & \ \Phi'^2 + \left[M^2 - \left(\epsilon - \frac{\alpha}{2}\omega\right)^2 + \frac{2}{r^2}\left(1 + \frac{1}{2}R_\alpha\right)^2\right]\Phi^2 + M^{*2}\left[\Psi_1^2 + \frac{1}{2}\Psi_2^2 - \Psi_0^2\right] \\ & + \frac{1}{2}\left[\Psi_2' - \frac{1}{r}\Psi_2\right]^2 + \frac{1}{r}R_\alpha \Psi_1 \Psi_2' + \frac{1}{r^2}R_\alpha(\Psi_1 + \Psi_2)\Psi_2 + \frac{1}{2r^2}R_\alpha^2\left(\Psi_1^2 + \frac{1}{2}\Psi_2^2\right) \\ & - \left[\Psi_0' - \left(\epsilon - \frac{\alpha}{2}\omega\right)\Psi_1\right]^2 - \frac{1}{2}\left[\frac{R_\alpha}{r}\Psi_0 + \left(\epsilon - \frac{\alpha}{2}\omega\right)\Psi_2\right]^2 \\ & - d\Bigg\{\frac{2}{r}\sin F\left[\Psi_2 \Psi_0' - \frac{R_\alpha}{r}\Psi_0\Psi_1 - \left(\epsilon - \frac{\alpha}{2}\omega\right)\Psi_1\Psi_2\right] \\ & + \frac{F'}{r}\left[\frac{r}{2}\left(\epsilon - \frac{\alpha}{2}\omega\right)\Psi_2^2 - (1-\cos F)\Psi_0\Psi_2\right]\Bigg\} + 2Md\left[F'\Psi_1 - \frac{\sin F}{r}\Psi_2\right]\Phi \\ & + \frac{2\sqrt{2}cM}{gm_V}\left[2\omega'\Psi_0\Psi_1 - \frac{2G'}{r}\Psi_1\Psi_2 + \frac{G}{2r^2}(G+2)\Psi_2^2\right] \\ & + \frac{4\sqrt{2}c}{gm_V}\Bigg\{\frac{1}{r^2}\left(\epsilon - \frac{\alpha}{2}\omega\right)[G(G+2)\Psi_1 - rG'\Psi_2]\Phi - \frac{\omega'}{r}\left[1 + \frac{R_\alpha}{2}\right]\Psi_2 \\ & + \frac{1}{r^2}[G(G+2)\Phi' + G'(2+R_\alpha)\Phi]\Psi_0\Bigg\}.\end{aligned} \quad (A.3)$$

Also for the P wave we display the leading term in the limit $M = M^* \to \infty$

$$\begin{aligned}\mathcal{L}_H^{(P)} \to & \left[M^2 - \left(\epsilon - \frac{\alpha}{2}\omega\right)^2\right]\left[\Phi^2 + \Psi_1^2 + \frac{1}{2}\Psi_2^2\right] + 2Md\left[F'\Psi_1 - \frac{\sin F}{r}\Psi_2\right]\Phi \\ & + d\left(\epsilon - \frac{\alpha}{2}\omega\right)\left[\frac{2}{r}\sin F \Psi_1 - \frac{F'}{r}\Psi_2\right]\Psi_2 + \frac{2\sqrt{2}cM}{gm_V}\left[\frac{2G'}{r}\Psi_1 - \frac{G}{2r^2}(G+2)\Psi_2\right]\Psi_2 \\ & + \frac{4\sqrt{2}c}{gm_V r^2}\left(\epsilon - \frac{\alpha}{2}\omega\right)[G(G+2)\Psi_1 - rG'\Psi_2]\Phi.\end{aligned} \quad (A.4)$$

Again the expansion (3.1) provides the results for the bound (anti–) mesons in the P wave channel, which are discussed in chapter 3.

The Euler–Lagrange equations associated with (A.1) and (A.3) are integrated using the method described in appendix A for ref. [12]. Technically a problem arises because the equation of motion for $\Psi_1$ is a constraint rather than a second order differential equation as in the case of the other fields. Hence $\Psi_1$ is not a dynamical degree of freedom and has to be eliminated in terms of the other fields. This can be achieved by formally writing

$$\mathcal{L} = \frac{1}{2}a_i\zeta_i'^2 + b_{ij}\zeta_i\zeta_j' + \frac{1}{2}c_{ij}\zeta_i\zeta_j + \frac{1}{2}f_1\Psi_1^2 + (g_i\zeta_i + h_i\zeta_i')\Psi_1 \tag{A.5}$$

where $\zeta_i$ refer to any of $\Phi, \Psi_0$ and $\Psi_2$. The coefficient functions $a_i$ etc. have to be identified from (A.1) and (A.3). Note that there are no terms linear in the meson fields. This is in contrast to the exploration of the induced kaon (and $K^*$) components of ref. [12]. The elimination of $\Psi_1$ yields

$$\mathcal{L} = \frac{1}{2}\tilde{A}_{ij}\zeta_i'\zeta_j' + \tilde{B}_{ij}\zeta_i\zeta_j' + \frac{1}{2}\tilde{C}_{ij}\zeta_i\zeta_j, \tag{A.6}$$

wherein

$$\tilde{A}_{ij} = a_i\delta_{ij} - \frac{h_i h_j}{f_1}, \quad \tilde{B}_{ij} = b_{ij} - \frac{h_i g_j}{f_1} \quad \text{and} \quad \tilde{C}_{ij} = c_{ij} - \frac{g_i g_j}{f_1}. \tag{A.7}$$

This equation also shows that singularities may appear when $f_1 = 0$, $cf$ Sect. 6. Noting now that

$$\tilde{A}_{ij}^{-1} = \frac{1}{a_i}\delta_{ij} + \left[f_1 - \sum_{l=1}^{3}\frac{h_l h_l}{a_l}\right]^{-1}\frac{h_i}{a_i}\frac{h_j}{a_j} \tag{A.8}$$

the differential equations resulting from (A.6) may be integrated by standard numerical techniques. In general no regular solution exists for an arbitrary value of the energy $\epsilon$. For the purpose of obtaining the regular solution $\epsilon$ is treated as a parameter. The value yielding this solution is identified as the bound state energy.

Finally we would like to incorporate the light vector mesons into the approximate bound state equation studied in ref.[9]. This bound state equation is derived in two steps. First the equations of motion for $\Phi$ are approximated by the leading contributions in the limit $M = M^* \to \infty$. In the second step the heavy limit relations (3.4) and (3.3) are substituted into these equations. This procedure results in

$$\Phi'' = -\frac{2}{r}\Phi' + \left[M^2 - \left(\epsilon - \frac{\alpha}{2}\omega\right)^2 + \frac{1}{2r^2}(l_{\text{eff}}(l_{\text{eff}}+1) + R_\alpha)^2\right]\Phi$$
$$-2Md\left[F' - \frac{2}{r}\sin F\right]\Phi - \frac{2c}{gm_V}\left(\epsilon - \frac{\alpha}{2}\omega\right)\left[\frac{2G'}{r} - \frac{G}{r^2}(G+2)\right]\Phi. \tag{A.9}$$

Here $l_{\text{eff}} = 1, 0$ refers to the S– and P waves, respectively. Once again we remind the reader that the presence of the soliton exchanges the behaviors of the S– and P wave–functions near $r \approx 0$. It should also be noted that the expansions (3.1) lead in a straightforward manner to the binding energies (3.2) in the heavy limit. Similarly the penta quark bound state energy (3.10) is obtained by assuming the relations (3.11) in the second step of the above described procedure.

*Appendix B: Conventions*

In order to agree with the conventions of refs. [6, 9] we should make the following replacements for the quantities in the present paper:

$$\begin{aligned} f_\pi &\to F_\pi/\sqrt{2}, \\ \rho_\mu &\to \frac{1}{\sqrt{2}}\rho_\mu, \\ g &\to \sqrt{2}\tilde{g}, \\ G(r) &\to -G(r), \\ \omega(r) &\to -\sqrt{2}\tilde{g}\omega(r). \end{aligned} \tag{B.1}$$

*References*

[1] C. Callan and I. Klebanov, Nucl. Phys. **B262** (1985) 365;
C. Callan, K. Hornbostel, and I. Klebanov, Phys. Lett. **B202** (1988) 296;
I. Klebanov in *Hadrons and Hadronic Matter*, page 223, proceedings of the NATO Advanced Study Institute, Cargese, 1989, edited by D. Vautherin, J. Negele and F. Lenz (Plenum Press 1989).

[2] J. Balaizot, M. Rho, and N. Scoccola, Phys. Lett. **B209** (1988) 27;
N. Scoccola, H. Nadeau, M. A. Novak, and M. Rho, Phys. Lett. **B201** (1988) 425;
D. Kaplan and I. Klebanov, Nucl. Phys. **B335** (1990) 45
Y. Kondo, S. Saito, and T. Otofuji, Phys. Lett. **B256** (1991) 316;
M. Rho, D. O. Riska, and N. Scoccola, Z. Phys. **A341** (1992) 341;
H. Weigel, R. Alkofer, and H. Reinhardt, Nucl. Phys. **A576** (1994) 477.

[3] E. Eichten and F. Feinberg, Phys. Rev. **D23** (1981) 2724;
M. B. Voloshin and M. A. Shifman, Yad. Fiz. **45** (1987) 463 (Sov. J. Nucl. Phys. **45** (1987) 292);
N. Isgur and M. B. Wise, Phys. Lett. **B232** (1989) 113; **B237** (1990) 527;
H. Georgi, Phys. Lett. **B230** (1990) 447.

[4] E. Jenkins and A. V. Manohar, Phys. Lett. **B294** (1992) 173;
Z. Guralnik, M. Luke, and A. V. Manohar, Nucl. Phys. **B390** (1993) 474;
E. Jenkins, A. V. Manohar, and M. Wise, Nucl. Phys. **B396** (1993) 27, 38.

[5] M. Rho, in *Baryons as Skyrme Solitons*, World Scientific, 1994, edited by G. Holzwarth;
D. P. Min, Y. Oh, B. Y Park, and M. Rho, *Soliton structure of heavy baryons* Seoul

report no. SNUTP 92–78, hep-ph/9209275;
H. K. Lee, M. A. Novak, M. Rho, and I. Zahed, Ann. Phys. (N.Y.) **227** (1993) 175;
M. A. Novak, M. Rho, and I. Zahed, Phys. Lett. **B303** (1993) 130.
D. P. Min, Y. Oh, B. Y Park, and M. Rho, *Heavy–quark symmetry and Skyrmions* Seoul report no. SNUTP 94–117, hep-ph/9412302.

[6] K. S. Gupta, M. A. Momen, J. Schechter and A. Subbaraman, Phys. Rev. **D47** (1993) R4835;
M. A. Momen, J. Schechter and A. Subbaraman, Phys. Rev. **D49** (1994) 5970.

[7] Y. Oh, B. Y. Park and D. P Min, Phys. Rev. **D49** (1994) 4649;

[8] Y. Oh, B. Y. Park and D. P Min, Phys. Rev. **D50** (1994) 3350.

[9] J. Schechter and A. Subbaraman, "Excited Heavy baryons in the bound state picture", SU–4240–586, UCI–TR 94–38.

[10] Y. Oh and B. Y. Park, *Energy levels of the soliton–heavy–meson bound states*, hep-ph/9501356.

[11] N. W. Park and H. Weigel, Phys. Lett. **B268** (1991) 155;

[12] N. W. Park and H. Weigel, Nucl. Phys. **A541** (1992) 453.

[13] H. Albrecht et al. (ARGUS collaboration), Phys. Lett. **B317** (1993) 227;
D. Acosta et al. (CLEO collaboration), CLEO conf. 93–7, contributed to the International Symposium on Lepton and Photon Interactions, Ithica (1993);
M. Battle et al. (CLEO collaboration), CLEO conf. 93–7, contributed to the International Symposium on Lepton and Photon Interactions, Ithica (1993);
P. L. Frabetti et al. (E687 collaboration), FERMILAB–Pub–93–32–E (1993).

[14] See also C.-K. Chow and M. B. Wise Phys. Rev. **D50**, 2135 (1994).

[15] G. Holzwarth, Nucl. Phys. **A572** (1994) 69.

[16] Ö. Kaymakcalan, S. Rajeev, and J. Schechter, Phys. Rev. **D30** (1984) 594.

[17] P. Jain, R. Johnson, Ulf-G. Meißner, N. W. Park and J. Schechter, Phys. Rev. **D37** (1988) 3252.

[18] Ulf.-G. Meißner, Phys. Rep. **161** (1988) 213;
B. Schwesinger, H. Weigel, G. Holzwarth, and A. Hayashi, Phys. Rep. **173** (1989) 173;
G. Holzwarth, editor, *Baryons as Skyrme Solitons*, World Scientific, 1994.

[19] J. Schechter and A. Subbaraman, Phys. Rev. **D48** (1993) 332.

[20] P. Jain, A. Momen and J. Schechter, "Heavy Meson Radiative Decays and Light Vector Meson Dominance", SU-4240-581.

[21] J. Schechter and H. Weigel, "Resolving Ordering Ambiguities in the Collective Quantization by Particle Conjugation Constraints", SU–4240–587, UNITU–THEP–22/1994.

[22] T. H. R. Skyrme, Proc. R. Soc **127** (1961) 260;
G. S. Adkins, C. R. Nappi, and E. Witten, Nucl. Phys. **B228** (1983) 552.

[23] M. A. Novak, M. Rho, and I. Zahed, Phys. Rev. **D48** (1993) 4370;
W. A. Bardeen and C. T. Hill, Phys. Rev. **D49** (1994) 409;
D. Ebert, T. Feldmann, R. Friedrich, and H. Reinhardt, Nucl. Phys. **B434** (1995) 619.